\documentclass[12pt]{article}

\renewcommand{\title}[1]{
\begin{center} \Large \bf #1 \end{center}
}

\renewcommand{\author}[2]{
 \begin{center} #1  \vspace{3mm} \\
  #2 \\
 \end{center}
\addvspace{\baselineskip}
}

\usepackage{amssymb}
\usepackage{amsmath}

\usepackage{amsthm}
\newtheorem{thm}{Theorem}[section]
\newtheorem{prop}[thm]{Proposition}

\theoremstyle{definition}

\theoremstyle{remark}

\makeatletter
\@addtoreset{equation}{section}

\makeatother

\begin{document}

\baselineskip 5mm

\title{Noncommutative Deformation of Instantons }

\author{${}^{\flat}$Yoshiaki Maeda, ${}^\dagger$Akifumi Sako }{
${}^{\flat}$ Department of Mathematics,
Faculty of Science and
 Technology, Keio University\\
3-14-1 Hiyoshi, Kohoku-ku, Yokohama 223-8522, Japan \\
${}^{\flat}$ Mathematical Research Centre, 
University of Warwick\\
Coventry, CV4 7AL , United Kingdom \\
${}^\dagger$ 
Department of General Education, Kushiro National College of Technology\\
Otanoshike-Nishi 2-32-1, Kushiro 084-0916, Japan }

\vspace{1cm}

\abstract{
We construct instanton solutions on noncommutative Euclidean 4-space
which are deformations of  instanton solutions on commutative
Euclidean 4-space.  We show that the instanton
numbers of these noncommutative instanton solutions  coincide
with the commutative solutions and conjecture that the instanton
number in ${\mathbb R}^4$ 
is preserved for general noncommutative deformations.
We also study noncommutative deformation of instanton solutions on a $T^4$
with twisted boundary conditions.
}

\section{Introduction}
\label{section1}

\newcommand{\bra}[2]{\left<#1,#2\right|}

Gauge theory originated in physics
as a convenient framework for electromagnetic fields and their generalizations to e.g. the Yang-Mills theories.
In mathematics, gauge theory has been highly developed 
to study the topology of 4-manifolds, with Donaldson's construction of
a new obstruction to the smoothability 
of 4-manifolds, which produced a series of examples of exotic differentiable
structures.  For these problems, it is important to study  the moduli spaces of
anti-self 
dual connections or instantons. These moduli spaces have an algebro-geometric
interpretation. In particular, anti-selfdual connections are classified by their 
instanton numbers.

Many authors have worked on extending
gauge theory to noncommutative
geometry.  Several authors have treated the ADHM construction
on noncommutative Euclidean 4-manifolds
\footnote{
Noncommutative instanton solutions were constructed with the ADHM method
in \cite{NCinstanton}. After that, many authors have constructed 
noncommutative instantons in the similar way.
See for example \cite{NCinstlecture} and their bibliography.}
and have shown that the
instanton number is given by an integer which does not depend
on the noncommutative parameter \cite{sako2,sako3,Furuuchi1,Furuuchi2,Tian}.  
We note that the relation  between these 
noncommutative instantons and  deformed solutions from the commutative ADHM 
construction \cite{ADHM} is unknown.
\footnote{ There are few noncommutative instanton solutions 
whose commutative limits are 
clarified, and they are constructed
without using the ADHM method \cite{CommLimi}.}

In the paper \cite{maeda_sako}, we constructed a noncommutative vortex
solution which is a deformation of   Taubes's vortex solution and
showed that its vortex number is undeformed, i. e. independent of the deformation parameter. It is therefore natural to 
construct a deformed instanton solution via the ADHM 
construction from the commutative one and to see
if the corresponding instanton number is deformed.  

In this paper, we construct a noncommutative formal instanton solution 
which is a deformation of the commutative instanton solution. Our
construction starts with a commutative instanton solution,
which is determined by its ADHM data, and then
solves the infinite systems of elliptic PDE equations 
with decay conditions term by term in the noncommutative parameter
$\hbar$.  We study the (noncommutative) instanton number for this 
noncommutative instanton solution and show that it is independent 
of  $\hbar$ (Theorem \ref{main}). 
This result supports our conjecture on the independence of 
the noncommutative instanton number for noncommutative 
${\mathbb R}^4$. 
We also study noncommutative deformations of 
instantons and their corresponding instanton numbers on $T^4$ with a twisted bundle.


\section{ Notations } \label{sect2}
Noncommutative Euclidean 4-space is given by the following commutation 
relations:
\begin{eqnarray}
[ x^{\mu} , x^{\nu} ]_\star = 
x^{\mu} \star x^{\nu} - x^{\nu} \star x^{\mu}= i \theta^{\mu \nu} ,
\ \mu , \nu = 1,2,3,4 \ ,
\end{eqnarray}
where $(\theta^{\mu \nu})$ is a real, $x$-independent, skew-symmetric matrix, called the
noncommutative parameters.
$\star$ is known as 
the Moyal product \cite{Moyal}.
The Moyal product (or star product) is defined on functions by
\begin{eqnarray}
 f(x)\star g(x)
  &:=&f(x)\exp\left(\frac{i}{2}\overleftarrow{\partial}_{\mu}
	      \theta^{\mu\nu}\overrightarrow{\partial}_{\nu}\right)g(x)
  \nonumber\\
 &=&f(x)g(x)+\sum_{n=1}^{\infty}\frac{1}{n!}f(x)
  \left(\frac{i}{2}\overleftarrow{\partial}_{\mu}
 \theta^{\mu\nu}\overrightarrow{\partial}_{\nu}\right)^ng(x)\;.
 \nonumber
\end{eqnarray}
Here $\overleftarrow{\partial}_{\mu}$ and $\overrightarrow{\partial}_{\nu}$ are  partial derivatives with
respect to $x^{\mu}$ for $f(x)$ and 
to $x^{\nu}$ for $g(x)$, respectively.

We define a  Lie algebra  by
\begin{eqnarray}
[ T_a , T_b ] = i f_{abc} T_c ,
\end{eqnarray}
where the generators $T_a$ are given by  Hermitian matices.
We consider a compact simply connected Lie group 
with this Lie algebra.
The covariant derivative
is defined by
\begin{eqnarray}
D_{\mu} := \partial_{\mu} + i A_{\mu} \ , \ \ 
A_{\mu}= A_{\mu}^a T_a .
\end{eqnarray}
The curvature two form $F$ is defined by
\begin{eqnarray}
F := \frac{1}{2} F_{\mu \nu} dx^{\mu} \wedge \star dx^{\nu} 
= dA+ A \wedge \star A 
\end{eqnarray}
where $\wedge \star $ is defined by 
\begin{eqnarray}
A\wedge \star A := \frac{1}{2}
(A_{\mu} \star A_{\nu} ) dx^{\mu} \wedge dx^{\nu}  .
\end{eqnarray}

To consider smooth noncommutative deformations,
we introduce a parameter $\hbar$ and a fixed constant
$\theta^{\mu \nu}_0 < \infty$ with
\begin{eqnarray}
\theta^{\mu \nu} = \hbar \theta^{\mu \nu}_0 .
\end{eqnarray}
We define the commutative limit by letting $\hbar \rightarrow 0$.

\section{Noncommutative Instantons}
Instanton solutions or anti-selfdual
connections  satisfy the (noncommutative) instanton equation
\begin{eqnarray}
F^+ = \frac{1}{2} (1 + *) F = 0 \ , \label{ASDEQ}
\end{eqnarray}
where $*$ is the Hodge star operator.
Formally we expand the connection as
\begin{eqnarray}
A_{\mu} = \sum_{l=0}^{\infty} A_{\mu}^{(l)} \hbar^{l}  .
\end{eqnarray}
Then,
\begin{eqnarray}
A_{\mu} \star A_{\nu} &=& \sum_{l,m,n=0}^{\infty}
\hbar^{l+m+n} \frac{1}{l~ !} A_{\mu}^{(m)} 
(\overleftrightarrow{\Delta} )^l A_{\mu}^{(n)} \\
\overleftrightarrow{\Delta} &\equiv& \frac{i}{2} \overleftarrow{\partial}_{\mu}
\theta^{\mu \nu}_0
\overrightarrow{\partial}_{\nu} . \nonumber
\end{eqnarray}
We introduce the selfdual projection operator $P$ by
\begin{eqnarray}
P := \frac{1+*}{2}  \ ; \ 
P_{\mu \nu , \rho \tau} 
=\frac{1}{2} (\delta_{\mu \rho} \delta_{\nu \tau} -
\delta_{\nu \rho} \delta_{\mu \tau} + \epsilon_{\mu \nu \rho \tau} ) .
\end{eqnarray}
Then the instanton equation is 
\begin{eqnarray} \label{ASDEQ_P}
P_{\mu \nu , \rho \tau} F^{\rho \tau} =0 . 
\end{eqnarray}
In the noncommutative case, the $l$-th order equation of (\ref{ASDEQ_P})
is given by
\begin{eqnarray}
&&P^{\mu \nu , \rho \tau} 
( \partial_{\rho} A_{\tau}^{(l)}- \partial_{\tau} A_{\rho}^{(l)}
+ i [A_{\rho}^{(0)} , A_{\tau}^{(l)} ] 
+ i [A_{\rho}^{(l)} , A_{\tau}^{(0)} ]
+ C_{\rho \tau}^{(l)} ) = 0
,  \label{k-th_order}\\
&&C_{\rho \tau}^{(l)} := 
\sum_{(p ;~ m,n) \in I(l)} 
\hbar^{p+m+n} \frac{1}{p~ !} \big( A_{\rho}^{(m)} 
(\overleftrightarrow{\Delta} )^p A_{\tau}^{(n)} -
A_{\tau}^{(m)} 
(\overleftrightarrow{\Delta} )^p A_{\rho}^{(n)}
\big)\  , \nonumber \\
&& I(l) \equiv 
\{( p ;~ m,n ) \in {\mathbb Z}^3 | p+m+n =l,~ p,m,n \ge 0 ,~
m\neq l , n\neq l \} . \nonumber
\end{eqnarray}
Note that the $0$-th order equation is the commutative instanton equation 
with solution  $A^{(0)}_{\mu}$
a commutative instanton.
The asymptotic behavior of commutative instanton $A^{(0)}_{\mu}$
is given by
\begin{eqnarray}
A^{(0)}_{\mu} = gdg^{-1}+O(|x|^{-2}), \ gdg^{-1} = O(|x|^{-1}) ,
\label{AsymptoticA}
\end{eqnarray}
where $g \in G$ and $G$ is a gauge group. 
We introduce covariant derivatives associated to the commutative instanton
connection by
\begin{eqnarray}
D^{(0)}_{\mu} f := \partial_{\mu} f + i [ A_{\mu}^{(0)} , f ] ,\ \
D_{A^{(0)}} f := d~ f + A^{(0)} \wedge f 
\end{eqnarray}
Using this, (\ref{k-th_order}) is given by
\begin{eqnarray}
&&P^{\mu \nu , \rho \tau} \big( 
D^{(0)}_{\rho} A_{\tau}^{(l)}- D^{(0)}_{\tau} A_{\rho}^{(l)}
+ C_{\rho \tau}^{(l)}
\big) =0 \nonumber \\
&& P (D_{A^{(0)}} A^{(l)} + C^{(l)} )=0 . \label{lthASD}
\end{eqnarray}

In the following, we fix 
a  commutative 
anti-selfdual connection $A^{(0)}$.
We impose the following 
condition for 
$A^{(l)} (l \ge 1 )$
\cite{Freed}
\begin{eqnarray}
A - A^{(0)}= D^{*}_{A^{(0)}} B \ , \ B \in \Omega^2_+ ,
\end{eqnarray}
where $D^*_{A^{(0)}}$ is defined by
\begin{eqnarray}
(D^{*}_{A^{(0)}} )^{\mu \nu}_\rho B_{\mu \nu}
&=& \delta^{\nu}_{\rho} \partial^{\mu}  B_{\mu \nu}
- \delta^{\mu}_{\rho} \partial^{\nu}  B_{\mu \nu}
+i \delta^{\nu}_{\rho} [ A^{\mu} , B_{\mu \nu} ]
-\delta^{\mu}_{\rho} [ A^{\nu} , B_{\mu \nu} ]
\nonumber \\
&=& \delta^{\nu}_{\rho} D^{(0)\mu} B_{\mu \nu} 
-\delta^{\mu}_{\rho} D^{(0)\nu} B_{\mu \nu} .
\end{eqnarray}
We expand $B$ in $\hbar$ as we did with $A$.
Then $A^{(l)}= D^{*}_{A^{(0)}} B^{(l)}$.
In this gauge, (\ref{lthASD}) is given by
\begin{eqnarray}
P D_{A^{(0)}} D^{*}_{A^{(0)}} B^{(l)}  + P C^{(l)} =0 . \label{lthASD2}
\end{eqnarray}
Using the fact that the $A^{(0)}$ is an anti-selfdual connection,
(\ref{lthASD2}) simplifies to
\begin{eqnarray} \label{main_equation1}
&&2  D_{(0)}^2 B^{(l) \mu \nu}
+  P^{\mu \nu , \rho \tau} C_{\rho \tau}^{(l)} =0 , 
\end{eqnarray}
where
$$ \ D_{(0)}^2 \equiv D_{A^{(0)}}^{\rho} D_{A^{(0)} \rho}\ .
$$

\section{ Green's Functions}

In this section, we derive some properties of the 
Green's function of $D_{(0)}^2$ in preparation for Theorem \ref{main}. 
To apply results from the ADHM construction, we restrict ourselves to $U(n)$
gauge theory. \\

We consider the Green's function for  $D_{(0)}^2$:
$$D_{(0)}^2 G_0 (x,y) = \delta(x-y) , $$
where $\delta(x-y)$ is a four dimensional delta function.
Here $D_{(0)}^2 \equiv D_{A^{(0)}}^{\rho} D_{A^{(0)}\rho}$, and 
this $A^{(0)}$ is an instanton 
in commutative ${\mathbb R}^4$.
Instantons in commutative ${\mathbb R}^4$ 
are given by the ADHM construction \cite{ADHM},
and arbitrary commutative instantons are in one-to-one 
correspondence with ADHM data. 
$G_0 (x,y)$ has been constructed in
\cite{Corrigan:1978xi} (see also \cite{Corrigan:1978ce,Christ:1978jy}):
\begin{eqnarray}
G_0 (x,y) = \frac{[v_1 (x) \otimes v_2(x) ]^{\dagger} (1 -{\mathfrak M})
[v_1 (y) \otimes v_2(y) ] }{4 \pi^2 (x-y)^2} . 
\end{eqnarray}
Here ${\mathfrak M}$ and $v_1 , v_2$ are determined by the ADHM data
and $v_i$ is a bounded function.
Using this Green's function, we solve the equation (\ref{main_equation1})
as
\begin{eqnarray}
B^{(l) \mu \nu}
= -\frac{1}{2} \int_{{\mathbb R}^4} 
G_0(x,y) P^{\mu \nu , \rho \tau} C_{\rho \tau}^{(l)} (y) d^4y 
\end{eqnarray}
and the noncommutative instanton $A= \sum A^{(l)}$ is given by
\begin{eqnarray}
A^{(l)} = D^*_{A^{(0)}} B^{(l)} .
\end{eqnarray}
The key fact used in the following proposition is that the asymptotic behavior of
Green's function of $D_{(0)}^2$ 
is given by 
\begin{eqnarray}
G_0 (x,y) = O(|x-y|^{-2}) \ , |x-y| >>1 \ . \label{green1}
\end{eqnarray}

We now list some features of Green's functions like $G_0$. 
\begin{prop} \label{theo1}
Let $G(x,y)$ be a Green's function on ${\mathbb R}^4$ 
written as
\begin{eqnarray}
G(x,y) = \frac{b(x,y)}{|x-y|^{2}} ,
\end{eqnarray}
where $b(x,y)$ is a bounded function.
Let $f(x)$ be a function such that $|f(x)|< \frac{C}{1+|x|^4}$
where $C$ is some constant. We define $F(x)$ by
\begin{eqnarray}
F(x):=\int_{{\mathbb R}^4} G(x,y) f(y) d^4y .
\end{eqnarray}
Then $F(x)=O(|x|^{-2})$ .
\end{prop}

Lemma (3.3.35) in \cite{D-K} contains a more general formula
but with a rougher estimate, so
we give a proof of this proposition.

\noindent
[Proof]
We introduce two balls whose radii are $\frac{1}{2}|x|$
with centers at the origin and $x$ in $\mathbb{R}^4$.
Let $B_0$ and $B_1$ denote these balls respectively,
and let $C$ be their complement. Then
\begin{eqnarray}
&&F(x)=\int_{{\mathbb R}^4} G(x,y) f(y) d^4y \nonumber \\
&&= \int_{B_0} G(x,y) f(y) d^4y 
+\int_{B_1} G(x,y) f(y) d^4y 
+\int_{C} G(x,y) f(y) d^4y . \label{1+2+3}
\end{eqnarray}
The first term is estimated as follows.
\begin{eqnarray}
\int_{B_0} G(x,y) f(y) d^4y 
&<&\int_{|y|\le \frac{1}{2}|x|}  \frac{C}{|x-y|^2 (1+|y|^4)}d^4y 
\nonumber \\
&\le& \int_{|y|\le \frac{1}{2}|x|}  \frac{C}{|x|^2 (1+|y|^4)}d^4y 
\hspace{1cm} (\mbox{because }\ |x-y|\ge \frac{1}{2}|x| ) \nonumber \\ 
&=&4\pi^2 C |x|^{-2} \int_0^{\frac{1}{2}|x|} \frac{r^3}{1+r^4} dr 
\nonumber \\
&=& \pi^2 C |x|^{-2} [\log (1+r^4) ]_0^{\frac{1}{2}|x|} =O(|x|^{-2})
\end{eqnarray}
The second term is estimated as follows.
\begin{eqnarray}
\int_{B_1} G(x,y) f(y) d^4y
&\le& 2\pi^2 \lim_{\epsilon \rightarrow +0}
\int_{\epsilon\le |y-x|\le \frac{1}{2}|x|}
\frac{r^3}{r^2 (|x|-r)^4} \nonumber \\
&=&-2\pi^2 \lim_{\epsilon \rightarrow +0} \left\{ 
\left[\frac{1}{6}t^{-3}\right]_{t=(|x|-\epsilon)^2}^{t=\frac{1}{4}|x|^2}
+|x|\left[ \frac{1}{3}(r-|x|)^{-3} \right]_{\epsilon}^{\frac{1}{2}|x|}
\right\} \nonumber \\
&=& O(|x|^{-2})
\end{eqnarray}
where $y= x+r \omega, r \in {\mathbb R_{\ge0}} , |\omega|=1$
and we use the fact that $|y|\ge ||x|-r|\ge |x|-r$. 
To estimate the last term in (\ref{1+2+3})
we introduce $D_1$ and $D_2$ by
\begin{eqnarray}
D_1 &:=& \{ y\ |\ |y|\ge \frac{1}{2}|x| , |y-x|\ge \frac{1}{2}|x|
, |y-x| \ge |y| \}
\\
D_2 &:=& \{ y\ |\ |y|\ge \frac{1}{2}|x| , |y-x|\ge \frac{1}{2}|x|
, |y-x| \le |y| \} . \nonumber
\end{eqnarray}
Then, 
\begin{eqnarray}
\int_{C} G(x,y) f(y) d^4y
&<& C(\int_{D_1}+\int_{D_1} ) \frac{1}{|x-y|^2 |y|^4} d^4y
\nonumber \\
&<& C\int_{D_1} \frac{d^4 y}{|y|^6} + 
C\int_{D_2} \frac{d^4 y}{|y-x|^6}
=O(|x|^{-2}).
\end{eqnarray}
\begin{flushright}
$\square $
\end{flushright}

We introduce the notation $O'(|x|^{-m})$
as in \cite{D-K}.
If $s$ is a function of ${\mathbb R}^4$
which is $O(|x|^{-m})$ as $|x|\rightarrow \infty$
and $|D_{(0)}^k s|=O(|x|^{-m-k})$, then 
we denote this natural growth condition by $s \in O'(|x|^{-m})$.

Examining the proof of Proposition \ref{theo1},
and keeping track of  estimates for higher derivatives,
we have the following (see Lemma 3.3.36 in \cite{D-K}).

\begin{prop}\label{D-Klemma3_3_36}
If $f(x) \in O(|x|^{-m})$ and $|D_{(0)}^2 f(x)|=O'(|x|^{-m-2})$,
then $f(x) \in O'(|x|^{-m})$.
\end{prop}

We apply these propositions to our case.

\begin{thm}
If $C^{(l)} \in O'(|x|^{-4}) $, then 
$|B^{(k)}|< O'(|x|^{-2}) $ 
\end{thm}
\noindent
[Proof]
{}It follows easily from the construction of  $G_0$ \cite{Corrigan:1978xi} and the choice of ADHM data that 
 our Green's function can be written as
\begin{eqnarray}
G_0 (x,y) = \frac{b(x,y)}{|x-y|^{2}} ,
\end{eqnarray}
where $b(x,y)$ is a bounded function. If 
$C_{\rho \tau}^{(l)}$ is  $O'(|x|^{-4})$
 Proposition \ref{theo1} implies that
 $B^{(k)}= O(|x|^{-2}) $.
It follows from Proposition \ref{D-Klemma3_3_36} that
$B^{(k)}= O'(|x|^{-2}) $

\begin{flushright}
$\square $
\end{flushright}

In our case,
$C^{(1)}_{\rho \tau} = O'(x^{-4})$ by (\ref{AsymptoticA}), and so
 $|B^{(1)}| < O'(|x|^{-2})$ from  Theorem 4.3 and 
$|A^{(1)}|<O'(|x|^{-3})$ as $A^{(l)}= D^{*}_{A^{(0)}} B^{(l)}$.
Repeating the argument $l$ times, we get
\begin{eqnarray}
|A^{(l)}|< O'(|x|^{-3+\epsilon}) ,\ \   \forall \epsilon > 0 \ .
\label{AsymptoticAL}
\end{eqnarray}

\section{Instanton Number}
The first  Pontrjagin number is
defined by 
\begin{eqnarray} \label{instnumber}
I_{\hbar}:= \frac{1}{8\pi^2} \int tr\ F \wedge\star F .
\end{eqnarray}
We rewrite (\ref{instnumber}) as
\begin{eqnarray}
\frac{1}{8\pi^2} \int tr F \wedge\star F =
\frac{1}{8\pi^2} \int tr\ d(A \wedge\star d A +
\frac{2}{3} A \wedge\star A \wedge\star A+ )
+\frac{1}{8\pi^2} \int tr P_{\star}
\end{eqnarray}
where
\begin{eqnarray}
P_{\star} = \frac{1}{3} \left\{ 
F \wedge\star A \wedge\star A + 2 A \wedge\star F \wedge\star A
+ A \wedge\star A \wedge\star F + A \wedge\star A \wedge\star A \wedge\star A
\right\} . \label{P_star}
\end{eqnarray}
$\int tr P_{\star} $ is $0$ in the commutative limit,
but does not vanish in noncommutative space.
The cyclic symmetry of trace is broken by
the noncommutative deformation.\\

{}The trace of the first three terms in (\ref{P_star}) equals 
\begin{eqnarray}
&&tr \{ F \wedge\star A \wedge\star A + 2 A \wedge\star F \wedge\star A
+ A \wedge\star A \wedge\star F \} \label{leavingTerms} \\
&&= tr \sum_{k=1}^{\infty} \sum_{l=1}^2 \frac{i^k}{2^k k!}  
\theta^{\mu_1 \nu_1} \cdots \theta^{\mu_k \nu_k} \left\{
\big( \partial_{\mu_1} \cdots \partial_{\mu_1} P_l ) \wedge
\big( \partial_{\nu_1} \cdots \partial_{\nu_1}Q_l \big) \right.
\nonumber \\
&& \left. { } \hspace{25mm} 
+\big(\partial_{\mu_1} \cdots \partial_{\mu_k} P_l ) \wedge 
\big( \partial_{\nu_1} \cdots \partial_{\nu_k}Q_l \big) \right\} ,
\nonumber
\end{eqnarray}
where 
\begin{eqnarray}
P_1 = A \wedge\star F , \ Q_1 = A , \ 
P_2 = A , \ Q_2 = F \wedge\star A 
\nonumber
\end{eqnarray}
The trace of the last term in (\ref{P_star}) is 
\begin{eqnarray}
&&tr A \wedge\star A \wedge\star A \wedge\star A \label{leavingTerm2}
\\
&&= \frac{1}{2} tr \sum_{k=1}^{\infty}  \frac{i^k}{2^k k!}  
\theta^{\mu_1 \nu_1} \cdots \theta^{\mu_k \nu_k} \left\{
\big( \partial_{\mu_1} \cdots \partial_{\mu_k} P_3 ) \wedge 
\big( \partial_{\nu_1} \cdots \partial_{\nu_k}Q_3 \big) \right.
\nonumber \\
&& \left. { } \hspace{25mm} 
+\big(\partial_{\mu_1} \cdots \partial_{\mu_k} P_3 ) \wedge
\big( \partial_{\nu_1} \cdots \partial_{\nu_k}Q_3 \big) \right\} ,
\nonumber 
\end{eqnarray}
where
\begin{eqnarray}
P_3 = A \wedge\star A \wedge\star A , \ Q_3 = A . \ 
\nonumber
\end{eqnarray}

We discuss a more general case in the following.
Let $P$ and $Q$ be an $n$-form and a $(4-n)$-form $(n=0,\dots , 4)$, respectively,
and let $P \wedge Q$ be $O(\hbar^k)$.
Consider
\begin{eqnarray}
\int_{{\mathbb R}^d} tr ( P \wedge \star Q - (-1)^{n(4-n)} Q \wedge \star P ). 
\label{PQ0}
\end{eqnarray}
Note that (\ref{leavingTerms}) and (\ref{leavingTerm2}) are 
sums of the form (\ref{PQ0}). 
The lowest order term in $\hbar$
vanishes because of the cyclic symmetry of the trace,
i.e.
$
\int tr ( P \wedge Q - (-1)^{n(4-n)} Q \wedge P ) =0.
$
The term of order   $\hbar$ is given by
\begin{eqnarray}
&&\frac{i}{2} \int_{{\mathbb R}^4} tr \{\hbar \theta^{\mu \nu}_0 
(\partial_{\mu}P \wedge \partial_{\nu} Q) \} \label{PQ1}\\
&&= \frac{i}{2} \int_{{\mathbb R}^4} d^4x 
\hbar \theta^{\mu \nu}_0 (n! (4-n)! ) 
\epsilon^{\mu_1 \mu_2 \mu_3 \mu_4 }
tr \{
\partial_{\mu} P_{\mu_1 \dots \mu_n} \partial_{\nu} Q_{\mu_{n+1} \dots \mu_4}
\} \nonumber \\
&&= \frac{i}{2} \int_{{\mathbb R}^4} (n! (4-n)! )
\epsilon^{\mu_1 \mu_2 \mu_3 \mu_4 }
tr \{
\big( \frac{1}{4} \epsilon_{\mu \nu \rho \tau} \theta^{\rho \tau}
dx^{\mu} \wedge dx^{\nu} \big) \nonumber \\
&& \ \ \ \ \ \ \ \ \ 
\wedge
\big( \partial_{\sigma} P_{\mu_1 \dots \mu_n} \partial_{\eta} Q_{\mu_{n+1} \dots \mu_4}  dx^{\sigma} \wedge dx^{\eta}  \big) 
\} \nonumber \\
&&= \frac{i}{2} \int_{{\mathbb R}^4} (n! (4-n)! )
\epsilon^{\mu_1 \mu_2 \mu_3 \mu_4 }
tr \{ (* \theta ) \wedge 
d( P_{\mu_1 \dots \mu_n} d Q_{\mu_{n+1} \dots \mu_4} ) \}
\nonumber \\
&&= \frac{i}{2} \int_{{\mathbb R}^4} (n! (4-n)! )
\epsilon^{\mu_1 \mu_2 \mu_3 \mu_4 }
tr \ d \{ (* \theta ) \wedge( P_{\mu_1 \dots \mu_n} d Q_{\mu_{n+1} \dots \mu_4} ) \} \nonumber
\end{eqnarray}
where $* \theta = \epsilon_{\mu \nu \rho \tau} \theta^{\rho \tau} dx^{\mu} \wedge d x^{\nu} / 4$ .
These integrals are zero if
$P_{\mu_1 \dots \mu_n}
d Q_{\mu_{n+1} \dots \mu_4}$ is $O'(|x|^{-(4 -1+ \epsilon) }) \ (\epsilon > 0)$. 
Similarly,  higher order terms in $\hbar$ in (\ref{PQ0}) can be
written as total divergences and hence vanish under the decay hypothesis.
{}This fact and (\ref{AsymptoticAL}) imply that
$\int tr P_{\star} = 0$.

{}From the above discussion and (\ref{AsymptoticAL}),
\begin{eqnarray}
\frac{1}{8\pi^2} \int tr F \wedge\star F &=&
\frac{1}{8\pi^2} \int tr d(A \wedge\star d A +
\frac{2}{3} A \wedge\star A \wedge\star A+ )
+\frac{1}{8\pi^2} \int tr P_{\star} \nonumber \\
&=& \frac{1}{8\pi^2} \int tr F^{(0)} \wedge F^{(0)},
\end{eqnarray}
where $F^{(0)}$ is the curvature two form of $A^{(0)}$. Thus
the instanton number is not deformed under 
noncommutative deformation. \\

Summarizing the above discussions, we get following theorems.
\begin{thm} \label{main}
Let $A^{(0)}_{\mu}$ be a commutative instanton solution in
${\mathbb R}^4$ given by the ADHM construction.
There exists a formal noncommutative instanton solution 
$A_{\mu}= 
\sum_{l=0}^{\infty} A_{\mu}^{(l)} \hbar^l$
such that the instanton number $I_{\hbar}$ defined by (\ref{instnumber})
is independent of the noncommutative parameter $\hbar$ . 
\end{thm}

\section{Instanton Numbers on Noncommutative Torus}
In the proofs in the previous sections, the key point is that 
the volume of the space is infinite.
Therefore it is natural to expect that instanton number depends
on the noncommutative parameter for noncommutative 
deformations of a finite volume space.
To study this phenomena, in this section we consider  noncommutative deformation 
of  instantons on  $T^4$.

We first consider  instantons with twisted boundary condition
on a commutative $T^4$ (see \cite{hamanaka,ho,morariu,ganor}).
The twisted boundary conditions for covariant derivatives $D_{\mu}(x)$ 
are given by
\begin{eqnarray}
D_{\mu}(x_{\nu} + 2\pi , x_{\rho} {\scriptsize \mbox{$(\rho \neq \nu )$}} ) = 
\Omega_{\nu}(x) D_{\mu} (x_{\nu} , x_{\rho} 
{\scriptsize \mbox{$(\rho \neq \nu )$}} ) 
\Omega_{\nu}^{\dagger}(x) .
\end{eqnarray}
For the simplicity, we chose $\Omega_{\nu}(x)$ by
\begin{eqnarray}
\Omega_1 (x_2) = e^{i\frac{m}{n}x_2} U_1, &&
\Omega_2 (x_1) = V_1, \nonumber \\
\Omega_3 (x_4) = e^{i\frac{m}{n}x_4} U_2, &&
\Omega_2 (x_1) = V_2,  \nonumber
\end{eqnarray}
for unitary matrices $U_i, V_i$ satisfying
\begin{eqnarray}
U_i V_j = \big( 1- \delta_{ij}(1-e^{-2\pi i \frac{m}{n}} ) \big)
 V_j U_i , \nonumber \\
U_i U_j = U_j U_i , \ V_i V_j = V_j V_i  \ ( i,j = 1,2 ) . \nonumber
\end{eqnarray}

These $\Omega_{\mu}$ satisfy the consistency conditions,
\begin{eqnarray} \label{consist}
\Omega_1 (x_2 + 2\pi ) \Omega_2 (x_1) =
 \Omega_2 (x_1 + 2\pi) \Omega_1 (x_2) , \nonumber \\
\Omega_3 (x_4 + 2\pi ) \Omega_4 (x_3) =
 \Omega_4 (x_3 + 2\pi) \Omega_3 (x_4) .
\end{eqnarray}

In \cite{hamanaka}, a $k^2$ instanton solution with 
these twisted boundary conditions
for $U(N^2)$ gauge theory 
is given by
\begin{eqnarray}
D_{1}= \partial_1 ,&& D_2 = \partial_2  + 
\frac{1}{2} \frac{k}{N} (x_1 {\mathbf 1}_N) \otimes {\mathbf 1}_N , 
\label{hamanaka} \\
D_3= \partial_3 ,&& D_4 = \partial_4  -
\frac{1}{2} \frac{k}{N} (x_3 {\mathbf 1}_N) \otimes {\mathbf 1}_N ,
\nonumber
\end{eqnarray}
where ${\mathbf 1}_N$ is the identity matrix of degree $N$.
These covariant derivatives are valid operators 
under the consistency conditions.
For this connection, 
\begin{eqnarray}
F_{12}=-F_{34}= -\frac{i}{2\pi} \frac{k}{N} {\mathbf 1}_N \otimes {\mathbf 1}_N ,\
F_{13}=F_{24}=F_{14}=F_{23}=0 ,
\end{eqnarray}
which obviously satisfies the instanton equation.
The instanton number  is given by $k^2$.

Let us deform this solution to a noncommutative instanton.
For simplicity, we chose the commutation relations as
\begin{eqnarray}
[ x_1 , x_2 ]_{\star} = 2\pi i \theta ,\ \ [ x_3 , x_4 ]_{\star} = 2 \pi i \theta,
\end{eqnarray}
with all other commutators zero.
After this noncommutative deformation,  
\begin{eqnarray}
D_{1}= \partial_1 ,\ D_2 = \partial_2  + 
f (x_1 {\mathbf 1}_N) \otimes {\mathbf 1}_N , 
\label{hamanaka} \\
D_3= \partial_3 ,\ D_4 = \partial_4  - 
f (x_3 {\mathbf 1}_N) \otimes {\mathbf 1}_N \nonumber
\end{eqnarray}
still satisfy the noncommutative instanton equation 
for an arbitrary constant $f$.
But the consistency conditions (\ref{consist})
restrict $f$ to 
$$f= \frac{k}{2\pi (N- k \theta )} . $$
Thus, covariant derivatives can be deformed smoothly
from those of the  commutative torus.
The instanton number is also deformed to
\begin{eqnarray}
\frac{1}{8 \pi^2} \int_{T^4} tr \ F \wedge \star F = 
\frac{k^2 N^2}{(N- k \theta )^2} . 
\end{eqnarray}

{}From this observation and  Theorem {\ref{main}}, a new question arises:
``Which instantons 
preserve their instanton number under 
noncommutative deformation?"  This question is left as an open problem.\\

\noindent
{\bf Acknowledgement}\\
Y.M and A.S are supported by KAKENHI No.18204006
(Grant-in-Aid for Scientific Research (A))
 and No.20740049
 (Grant-in-Aid for Young Scientists (B)), respectively.
We would like to thank Steven Rosenberg for his helpful suggestions.



\end{document}